\documentclass[smallabstract,smallcaptions]{dccpaper}

\usepackage{subfig}
\usepackage{epsfig}
\usepackage{citesort}
\usepackage{amsmath}
\usepackage{amssymb}
\usepackage{comment}
\usepackage{diagbox}
\usepackage{color}
\usepackage{multirow}
\usepackage{url}
\usepackage[ruled,vlined,linesnumbered]{algorithm2e}
\usepackage{booktabs}
\usepackage{multirow}
\usepackage{subfig} 
\usepackage{rotating}

\usepackage[utf8]{inputenc}
\usepackage{epstopdf}
\usepackage{tikz}

\newlength{\figurewidth}
\newlength{\smallfigurewidth}

\newcommand*\circled[1]{\tikz[baseline=(char.base)]{
		\node[shape=circle,draw,inner sep=.75pt] (char) {#1};}}

\newcommand{\no}[1]{}

\newcommand{\vtvdc}{{\em v2vdc}}
\newcommand{\vtvdch}{{\em v2vdc}$_H$}

\newcommand{\etdc}{{\em etdc}}
\newcommand{\scdc}{{\em scdc}}
\newcommand{\detdc}{{\em detdc}}
\newcommand{\dscdc}{{\em dscdc}}
\newcommand{\dph}{{\em dph}}
\newcommand{\dvtv}{{\em D-v2v}}

\newcommand{\thuffman}{{\em TH}}
\newcommand{\phuffman}{{\em PH}}

\newcommand{\bzip}{{\em bzip2}}
\newcommand{\repair}{{\em re-pair}}

\newcommand{\psevenzip}{{\em p7zip}}
\newcommand{\lzma}{{\em lzma}}

\begin{document}

\title
{{\large\textbf{Dv2v: A Dynamic Variable-to-Variable Compressor}}
	\footnote{$^{\S}$ {\em Funded in part by 
		European Union's Horizon 2020 research and innovation programme
		under the Marie Sklodowska-Curie grant agreement No 690941 (project BIRDS). 
		G.N. is funded 
		  by the Millennium Institute for Foundational Research on Data (IMFD), Chile. 
		The Spanish group is funded 
		  by Xunta de Galicia/FEDER-UE [CSI: ED431G/01 and GRC: ED431C 2017/58]; 
  		  by MINECO-AEI/FEDER-UE [ETOME-RDFD3: TIN2015-69951-R, Datos 4.0: TIN2016-78011-C4-1-R];  		  
		  by MINECO-CDTI/FEDER-UE [INNTERCONECTA: uForest ITC-20161074];
          by FPI Program [BES-2017-081390] (T.V.R.); and by FPU Program [FPU16/02914] (A.G.B.).
	}
}}

\author{%
	Nieves R. Brisaboa$^{\ast}$, Antonio Fari{\~{n}}a$^{\ast}$,  Adrián Gómez-Brandón$^{\ast}$, \\ Gonzalo Navarro$^{\dag}$, and Tirso V. Rodeiro$^{\ast}$\\[0.5em]
	{\small\begin{minipage}{\linewidth}\begin{center}
				\begin{tabular}{ccc}
					$^{\ast}$Universidade da Coru\~na & \hspace*{0.5in} & $^{\dag}$University of Chile \\
					Facultade de Inform\'atica, CITIC && Millennium Institute for Foundational \\
					A Coruña, Spain &&  Research on Data (IMFD),\\
					\url{brisaboa}@udc.es,\url{fari@udc.es},&& Department of Computer Science \\
					\url{adrian.gbrandon@udc.es}, &&  Santiago, Chile\\
					\url{tirso.varela.rodeiro@udc.es} && \url{gnavarro@dcc.uchile.cl}\\
				\end{tabular}
	\end{center}\end{minipage}}
}

\maketitle
\thispagestyle{empty}

%
%
%
%
%
%

\begin{abstract}
	%
	We present \dvtv, a new dynamic (one-pass) variable-to-variable compressor. 
	Variable-to-variable compression aims at using a modeler that gathers variable-length  input symbols and a variable-length statistical coder that assigns shorter codewords to the more frequent symbols. 
	In \dvtv, we process the input text word-wise to gather variable-length symbols that  can be either terminals (new words) or non-terminals, subsequences of words seen before in the input text. Those input symbols are set in a vocabulary that is kept sorted by frequency. Therefore, those symbols can be easily encoded with dense codes. Our \dvtv\ permits real-time transmission of data, i.e. compression/transmission can begin as soon as data become available. 
	Our experiments show that \dvtv\ is able to overcome the compression ratios of the v2vDC, the state-of-the-art  semi-static variable-to-variable compressor, and to almost reach \psevenzip\ values. It also draws a competitive performance at both compression and decompression. 
\end{abstract}

\section{Introduction}

Text compression has gained relevance in the last decades along with the growth of text databases. It permitted not only to drastically reduce the storage needs of those data and the time needed to transmit them through a network, but also to handle them efficiently in compressed form. 

The first compressors based on Huffman coding \cite{Huffman:1952} using character-oriented modeling obtained rather poor compression ratios on text collections (compression around $60$\%). However, when Huffman coding was coupled with a word-based modeler during the 80s \cite{moffat:1989} the compression ratio obtained by those semi-static compressors was close to $25$\% when applied to English texts, and they set the basis to build modern text retrieval systems over them \cite{Witten:1999}. This boosted the interest of new compressors not only yielding fast decoding/retrieval but also  allowing queries to be performed in compressed form. At the end of the 90s, {\em Plain Huffman }(\phuffman)  and {\em Tagged Huffman }(\thuffman) \cite{Moura:1998,Navarro:2000} replaced the bit-oriented Huffman by byte-oriented Huffman to speed up decoding at the cost of loosing compression effectiveness (now around $30$\%). In addition, {\em TH} reserved the first bit of each byte to gain synchronization capabilities. Compression ratios worsened to  around $34$\% but random decompression and fast Boyer-Moore type searches became possible. In the same line, the use of {\em dense codes} \cite{BFNPir07} allowed {\em} End-Tagged Dense Code (\etdc) and {\em (s,c) Dense Code} (\scdc) to not only retain the same capabilities of {\thuffman} but also improve its compression ratios, which became very close to those of \phuffman, and a simpler coding scheme that does not depend on the Huffman tree. Indeed, assuming we have $n$ source symbols $s_i$ ($0\leq i<n$) with decreasing probabilities, the codeword $c_i$ corresponding to the $i$-th symbol can be obtained as $c_i\leftarrow encode(i)$, and the rank $i$ corresponding to $c_i$ can be obtained as $i\leftarrow decode(c_i)$. Both {\em encode} and {\em decode} algorithms perform in $O(|c_i|)$ time \cite{BFNP:SPE2008}. 

Unfortunately, since \phuffman\ is the optimal  word-based $256$-ary zero-order compressor, all those efficient and searchable word-based compression techniques, could never reach the compression of the strongest compressors (e.g. \psevenzip). This motivated the creation of \vtvdc, the first word-based {\em variable-to-variable}\footnote{Variable-to-variable compression aims at using a modeler that gathers variable-length symbols and a variable-length statistical coder that assigns shorter codewords to the more frequent symbols.} compressor \cite{BFLNL:dcc10}. \vtvdc\ obtained similar compression ratios to those of \psevenzip\ in English texts by parsing the text into both words and phrases (sequences of words) and then assigning codewords to them by using dense codes. Finally, the original words/phrases were replaced to create the compressed file. \vtvdc\ improved the compression ratio of \etdc\ by around $8\!-\!10$ percentage points 
and besides it  produced a still searchable compressed text.

\medskip
Data transmission is another scenario where compression is of special interest. In some cases, the whole data is available and can be compressed with the most powerful compressors. However, there are scenarios where dynamic (one-pass) real time compression becomes necessary.  That is, the compressor/sender must be capable of compressing the symbols on the fly as they arrive without the need of having the whole data before starting its compression/transmission. This could be the case of sensor data transmitted to a server, a digital library streaming a book to a electronic reader one page at a time, HTTP pages sent by a server during a HTTP session, etc. 
Even there exists powerful one-pass adaptive compressors such as those coupling arithmetic coding with k-order PPM-modeling \cite{Moffat:2002:CCA:560324}, or those derived from the Lempel-Ziv family \cite{zl77,zl78}, they do not match real-time requirements.
Yet, we can find in the literature versions of dynamic character-based Huffman compressors \cite{fal73,gal78} and also  word-based compressors such as the {\em Dynamic} \phuffman\ (\dph) \cite{phdfari2005}, or the {Dynamic} \etdc\  (\detdc) \cite{phdfari2005,BFNP:SPE2008}. The later takes advantage of the simple on-the-fly {\em encode} and {\em decode} algorithms from \etdc\  and permits both sender/compressor and receiver/decompressor to remain synchronized by simply keeping the same vocabulary of words sorted by frequency. Basically, assuming that, at a given moment, the vocabulary of the sender contains $n$ words, when the sender inputs the next word $w$ it could find it in its vocabulary at position $i$, so it simply sends $c_i \leftarrow encode(i)$ to the receiver. Otherwise, if $w$ is a new word, it sends $c_{n} \leftarrow encode(n)$ (used as an escape codeword) followed by $w$ in plain form. In any case, the encoder increases the frequency counter $f$ of $w$ to $f+1$ and runs a simple {\em update} algorithm that swaps $w$ with the first word that has frequency equal to $f$. This {\em update} algorithm keeps the vocabulary of words sorted by frequency and runs in $O(1)$ time. The receiver is also very simple. 
%
%
It receives a codeword $c_i$ and runs $i\leftarrow decode(c_i)$. Then, if $i<n$ it has decoded the word $w_i$ at the $i$-th entry of the vocabulary. Otherwise, if $i=n$ it receives a new word in plain form and adds it  at the end of the vocabulary. Finally,  a similar {\em update} procedure to that of the sender is run to increase the frequency of that word and to keep the vocabulary sorted. A variant based on \scdc\ (\dscdc) is also available \cite{BFNP:SPE2008}. Finally, more recent {\em lightweight} versions of \detdc\ and \dscdc, using asymmetric compression/decompression procedures to reduce the work done at decompression, were also created \cite{BFNP.tois10}. \detdc\ 
displayed similar compression ratios to those of the semi-static \etdc\ 
(around $33$\%) while yielding fast compression and decompression. Yet, as for \etdc, its compression effectiveness is far from the stronger variable-to-variable counterpart \vtvdc. 

\medskip
In this work, we 
create a dynamic (one-pass) variable-to-variable variant of \detdc\ 
named \dvtv. We  follow the same ideas from \detdc\ to keep both sender and decoder synchronized. The variable-length symbols in our vocabulary will be arbitrary length sequences of words that can be either single-words (terminals) or repeating pairs of symbols (that again can be terminals or not) that have occurred previously in the input sequence. We use a sort of Patricia tree \cite{DBLP:journals/jacm/Morrison68} to efficiently handle the subsequences seen before in the input sequence.
Note that our ability to choose {\em good/relevant} variable-length sequence of words will determine the success of our new compressor. However, finding the smallest grammar for a text is 
a NP-complete problem \cite{CLLPPSS05}. To overcome this, several heuristics exist: {\em lz78} \cite{zl78}  looks for existing substrings in the already processed sequence; {\em repair}  replaces pairs of repeating symbols recursively \cite{larsson:1999}; {\em Sequitur} \cite{NMWM04} replaces a pair of repeating phrases in the processed sequence by a new phrase; or others such as \cite{apostolico:2000} where the grammar induced ensures that non-terminals do only contain terminals. 
In \dvtv, our strategy to gather input symbols representing variable-length sequence of words is similar to that of {\em Sequitur}, whereas the semi-static \vtvdc\ followed the approach from \cite{apostolico:2000} supported by a {\em suffix array} \cite{MM93} of the text and the corresponding {\em longest common prefix} structure. 
We will show that our simpler procedure also 
yields competitive compression values.

\section{Our proposal: {\dvtv} } \label{sec:proposal}
As we have explained, \dvtv\ is a dynamic (one-pass) compressor. 
\dvtv\ processes the input text and gathers symbols that represent sequences with  a variable number of words. We use a sort of trie to help the parser to detect sequences of words that appeared before.  We keep those symbols sorted by frequency. In this way, we can use the \etdc\ encoder to encode them directly from their positions in the vocabulary. The decompressor/receiver is simpler because it only has to decode the received codewords and to keep the table of symbols sorted by frequency (synchronized with the sender).

In the next sections we conceptually describe the \textit{parser} and the \textit{encoder} procedures of the sender/compressor component,  and also the \textit{decoder} procedure that is the core of the receiver/decompressor component.



\subsection{Parsing algorithm used by the sender}

Our parser scans the text and splits it into tokens/symbols of one or more words that can be:
\begin{itemize}
	\item \textit{terminal symbols}. Those representing just one word. They are created when a new word is parsed.
	\item \textit{non-terminal symbols}. Those composed by two different symbols, which can be terminals or non-terminals. Therefore, each non-terminal, represents at least a sequence of two words.
\end{itemize}

\begin{figure}
	\centering
	\includegraphics[width=0.5\textwidth]{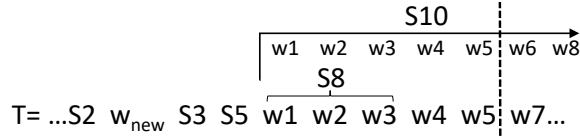}
	\caption{Non-terminal creation example.}
	\label{f:noTerminal}
	\vspace{-0.4cm}
\end{figure}

During the parsing, the sender reads the text one word at a time. If the next read word was not in our vocabulary, two symbols are created: {\em i)} a terminal symbol  $S_{new}$ which represents the new word (the sender will notify the receiver about this new word as we will show in the next section); and {\em ii)}  a non-terminal $S_{new+1}$ which appends $S_{new}$ to the previous sent symbol.
For example, in Figure~\ref{f:noTerminal}, after sending $S_2$ we read the new word $w_{new}$. Therefore, a terminal symbol $S_{new}$ is created for that word and then a non-terminal symbol $S_{new+1}$ is created for $S_2||S_{new}$.  

Otherwise, if the next read word is a prefix of any symbol from the vocabulary, we store such word in $RS$ (read sequence). We keep reading the text word by word and append those words to $RS$ until $RS$ becomes an unknown sequence. At this moment, we send the symbol which corresponds to the longest known prefix of $RS$. Then, a new non-terminal symbol containing the current sent symbol and the previous one is created. 
In the example of Figure~\ref{f:noTerminal}, let us assume that  $S_8$ is a non-terminal that expands into the sequence $w_1w_2w_3$, and $S_{10}$ produces $w_1w_2w_3w_4w_5w_6$, where $S_{10}$ is the unique non-terminal symbol starting by $w_1w_2w_3w_4w_5$. After sending the symbol $S_5$, we are at $w_1$ and we read the next words  $w_1 \dots w_5$ one word at a time. We keep reading words until we reach $w_7$. At that moment, $RS\leftarrow \underline{w_1w_2w_3w_4w_5}w_7$ is not a prefix of the sequence in $S_{10}$ and we stop processing the text. Note that, since the symbol containing the longest known prefix of $RS$ corresponds to $S_8$,   we send $S_8$  (the way to encode $S_8$ will be explained in detail in the next section)  and we create a new non-terminal symbol $S_{new}$ for $S_5||S_8$. We will continue parsing from $w_4$ on. 

In practice, we are using a set of known sequences $KS$ which stores every previously created terminal and non-terminal symbols. If we are sending the message ``\textit{the more I know about you the more I know about me}", at the beginning 
we have $KS=\emptyset$ and we  read the first word ``\textit{the}". Since $KS$ is empty, there is no sequence which starts with ``\textit{the}", thus we add it to $KS$, at position $i=|KS|=0$, the symbol $S_i =S_0=$ ``\textit{the}", and we send 
$S_0$ to the receiver.  Then we read ``\textit{more}", which is also a new word. Now we have to add both the new terminal symbol $S_1=$``\textit{more}" ($S_1$ is  also sent to the receiver) and the new non-terminal symbol $S_2=$ ``\textit{the more}" to $KS$.

After processing the word ``\textit{you}", $KS$ is composed by \{$S_0$:``\textit{the}", $S_1$:``\textit{more}", $S_2$:``\textit{the more}", $S_3$:``\textit{I}", $S_4$:``\textit{more I}", $S_5$:``\textit{know}", $S_6$:``\textit{I know}", $S_7$:``\textit{about}", $S_8$:``\textit{know about}", $S_9$:``\textit{you}", $S_{10}$:``\textit{about you}"\} and we continue reading ``\textit{the}". Since the current read sequence $RS=$``\textit{the}" exists in $KS$, we read the next word and append it to $RS$. Now, $RS=$``\textit{the more}" matches  the symbol $S_2$ stored in $KS$. In the next step, we update $RS$ to ``\textit{the more I}". Since that sequence is not included in $KS$, we send 
$S_2$ to the receiver and we create a new non-terminal 
that includes the previous and the current sent symbol: $S_{11}=S_9||S_2=$``\textit{you the more}".

We need a mechanism to check if $RS$ is within the set of known sequences $KS$ and to obtain its symbol identifier, i.e. its rank in $KS$. In order to perform those tasks efficiently we use a structure based on the Patricia-tree, where each branch represents a sequence, and all the sequences that start with the same prefix descend from the same node.\footnote{We implemented a bit-oriented trie where unary paths are stored in their parent node.} The last property is important, as it allows us to search incrementally for the longest sequence contained in $RS$. 
For example, after reading the second ``\textit{the}" ($RS=$``\textit{the}") we access to the trie of Figure~\ref{f:structures} and go through the branch labeled with ``\textit{the}" reaching the node-\circled{0}, which contains the identifier of  $S_0$. Then, we read ``\textit{more}" ($RS=$``\textit{the more}"), hence we descend from node-\circled{0} to node-\circled{2}, which contains the symbol $S_2$. Finally, we read ``\textit{I}" ($RS=$``\textit{the more I}"). Since we cannot descend from node-\circled{2}, the longest known sequence is ``\textit{the more}" and its symbol is $S_2$. 

\subsection{Encoding procedure }

Every parsed symbol  must be encoded and sent to the receiver. We encode them using \etdc\ dense codes. We need to keep track of the number of times each symbol was sent (frequency) because, following \etdc\ procedure, the codeword of a symbol depends only on its rank within the vocabulary sorted by frequency. Recall  \etdc\ assigns the shortest codewords to the most frequent symbols. Note that, each time a symbol is sent, its frequency is increased, and the codewords assigned to the symbols may change.
For each parsed symbol $S_i$ we send one codeword. In addition, when we send a terminal symbol  for the first time ($S_i = S_{|KS|})$, we send that codeword, which acts as an escape codeword,  followed by the word in plain format.

In order to encode the symbols, we use a \textit{codebook} where we store all the information required to compute the codeword of each symbol. Each entry in \textit{codebook} corresponds to a symbol $S_i$ and stores a tuple $\langle l, r, freq, voc\rangle$ as shown in Figure~\ref{f:structures}(left). $l$ and $r$ represent the sequence of each symbol. If the symbol is a non-terminal, $l$ and $r$ are pointers to the entries of \textit{codebook} where the left and right symbols of the non-terminal are stored. Otherwise, if the symbol is a terminal, $l$ stores the word itself and $r$ is set to -1. $freq$ stores the frequency of the symbol\footnote{Every non-terminal symbol is created with frequency 0.} and $voc$ holds the position of the symbol within the vocabulary sorted by frequency. The codeword $c_i$ corresponding to the symbol $S_i$ stored in the $i$-th entry of the codebook  is obtained as $c_i \leftarrow $\etdc$.encode(voc[i])$).

To keep the vocabulary sorted by frequency we use two arrays: \textit{pos} and \textit{top}.
Array \textit{pos} keeps the symbols sorted by frequency in decreasing order. Actually, $pos[i] = j$ indicates that the $i$-th most frequent symbol is stored in the $j$-th entry of the {\em codebook}. Consequently, note that all the symbols with the same frequency are pointed to from consecutive entries in \textit{pos}. Array \textit{top} contains a slot for each frequency value.  For every possible value of a frequency $f$, $top[f]=x$ means that the first symbol with frequency $f$ is at position $x$ in $pos$.  
For example, in Figure~\ref{f:structures}(left) the array \textit{top} indicates that the codewords of frequency 1 start at position 0 within \textit{pos}. We can observe that the gap between $top[0]$ and $top[1]$ is 6, thus $pos[0..5]$ point to the 6 entries within {\em codebook} that hold all the symbols with frequency 1.

With the help of the arrays \textit{pos} and \textit{top}, we can easily add the new symbols at the end of \textit{codebook}. Those arrays are also necessary to update the frequencies and positions in the vocabulary in $O(1)$ time without reordering the \textit{codebook}. In our example, after inserting ``\textit{you}", the table remains in the state of Figure~\ref{f:structures}(left). As we explained before, in the next step we send $S_2=$``\textit{the more}" which is the symbol at position 2. Therefore, we increase the frequency at $freq[2]$ to 1. We look for the position of the first symbol with frequency 0 by using $top[0]=6$. After that, we swap $voc[2]$ and $voc[6]$, so now $voc[2]=6$ and $voc[6]=9$. As we changed $voc$, we also have to update $pos$ accordingly. Therefore, we modify $pos[6]=2$ and $pos[9]=6$. Finally, as now the list of symbols with $f=1$ has been increased by one, the list of words with $f=0$ starts one position further, so we update $top[0]=7$.

\subsection{Receiver procedure }


The receiver works symmetrically to the sender. It decodes either a codeword corresponding to a known symbol or an escape codeword followed by a new word (terminal) in plain form. After decoding a symbol, we also add a new non-terminal composed of the last two decoded symbols to keep the codebook synchronized with the sender. This allows the receiver to rebuild the same model handled by the sender and to recover the original text. To carry this out the receiver also has a \textit{codebook} and an auxiliary \textit{top} array. The \textit{codebook} is composed of columns \textit{offset}, \textit{length}, and \textit{freq}. Each time we create a new symbol (i.e. we either received a new word or  we created a new non-terminal), we set in \textit{offset} a pointer to the position of the first occurrence of that symbol within the decompressed text. The length (in chars) of the text represented by such symbol is kept in \textit{length}. $freq$ stores the frequency of the symbol. 

\begin{figure}
	\centering
	\includegraphics[width=0.95\textwidth]{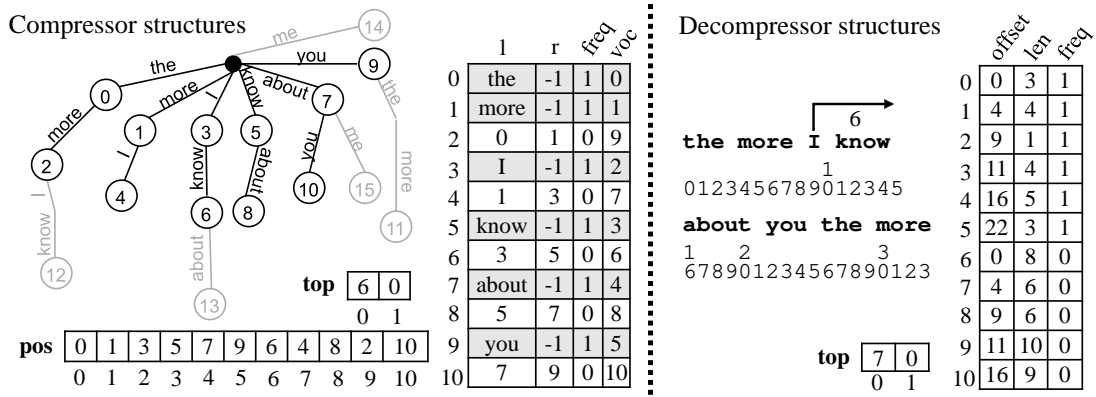}
	\caption{Strutures used in both compressor (left) and decompressor (right) when processing the sentence: ``{\tt the more I know about you the more I know about me}''. The black branches in the trie represent its stage after processing the word ``{\tt you}''.}
	\label{f:structures}
\end{figure}

In Figure~\ref{f:structures}(right) we can observe the state of the receiver after decompressing ``\textit{the more I know about you the more}". Now the sender transmits the symbol $S_8=$``\textit{I know}" encoded with \etdc. The receiver decodes the codeword into $8$. It accesses to the \textit{codebook} at position $8$ and retrieves \textit{offset}$[8]=9$ and $length[8]=6$, thus the decoder recovers the sequence ``\textit{I know}" from the decompressed text from position $9$ to $14$. Afterwards, we update the decompressed text to ``\textit{the more I know about you the more}"$||$``\textit{ I know}". Then,
we increase $freq[8]$, and we swap the rows in the \textit{codebook} at positions $8$ and $top[0]=7$ (recall $top[0]$ is the first row with frequency equals to $0$). Finally, since the first row with frequency equals to 0 is moved to the next position, we update $top[0]=8$.

%




\section{Experimental evaluation}
We performed experiments to compare the compression effectiveness as well as the performance at compression and decompression of \dvtv\ with those of \detdc\footnote{\url{http://vios.dc.fi.udc.es/codes}} and \vtvdc, which are respectively the previous dynamic word-based technique that makes up the basis of \dvtv, and the state-of-the-art when considering semi-static variable-to-variable compression based on dense codes. In the case of \vtvdc, we considered the two variants proposed in \cite{BFLNL:dcc10}, i.e. \vtvdc\ and \vtvdch. The former one uses a simpler heuristic to gather phrases, whereas the latter uses a more complex heuristic that yields better compression at the cost of increased compression time. Given that in \dvtv\ new words are sent in plain form, we included two variants of \vtvdc\ and \vtvdch\ that%
, as in \cite{BFLNL:dcc10}, respectively represent the words in the vocabulary in plain form or compressed with \lzma.
In addition, we have included some of the most well-known representatives from different families of compressors
: \psevenzip\ and \lzma,\footnote{\url{http://www.7-zip.org}} \bzip,\footnote{http://www.bzip.org} and  an implementation of \repair,\footnote{\url{http://raymondwan.people.ust.hk/en/restore.html}} coupled with a bit-oriented Huffman.\footnote{\url{https://people.eng.unimelb.edu.au/ammoffat/mr_coder/}}

We used three text datasets from {\sc trec-2} and {\sc trec-4} named
Ziff Data 1989-1990 (ZIFF), Congressional Record 1993 (CR) and Financial Times
1991 (FT91). In addition, we created a large dataset (ALL) including ZIFF and AP-newswire from {\sc trec-2}, as well as 
Financial Times 1991 to 1994 (FT91, FT92, FT93, and FT94) and ZIFF from {\sc trec-4}. We also included three highly repetitive text datasets: \textit{world\_leaders.txt} (WL), \textit{english.001.2.txt} (ENG) and \textit{einstein.en.txt} (EINS) from {\sc pizzachili}.\footnote{\url{http://pizzachili.dcc.uchile.cl}}

Our test machine is an Intel(R) Core(TM) i7-3820@3.60GHz CPU (4cores-8siblings) with 64GB of DDR3-1600Mhz. It runs Ubuntu 12.04.5 LTS (kernel 3.2.0-126-generic). We compiled with gcc 4.6.4 and optimizations \texttt{-O9}. Our time results measure {\sc cpu} user time.

\begin{table}[htbp]
	\centering
	\caption{Compression ratio (\%) with respect to the size of the plain text dataset.}
	\scriptsize	
	\setlength\tabcolsep{4.5pt} 
	\begin{tabular}{|r|r|rr|rr|r|r|r|r|r|r|r|}
		\hline
		\multirow{2}[4]{*}{} & \multicolumn{1}{c|}{\multirow{2}[4]{*}{Detdc}} & v2vdc & v2vdc$_H$ & v2vdc & v2vdc$_H$ & \multicolumn{1}{c|}{\multirow{2}[4]{*}{D-v2v}} & Repair & \multicolumn{2}{c|}{lzma} & \multicolumn{1}{c|}{\multirow{2}[4]{*}{p7zip}} & bzip2 & \textbf{Size Plain} \\
		\cline{3-6}\cline{8-10}\cline{12-12}          &       & \multicolumn{2}{c|}{\textit{lzma words}} & \multicolumn{2}{c|}{\textit{plain words}} &       & \textit{+sHuff} & \textit{def} & \textit{-9 -e} &       & \textit{def} & \textbf{(KB)} \\
		\hline
		\textbf{FT91} & 35,64 & 27,15 & 26,65 & 30,11 & 29,61 & 28,60 & 24,00 & 25,50 & 25,25 & 25,52 & 27,06 & \textbf{14.404} \\
		\textbf{CR} & 31,99 & 23,55 & 23,13 & 24,73 & 24,31 & 22,86 & 20,16 & 22,05 & 20,83 & 21,63 & 24,14 & \textbf{49.888} \\
		\textbf{ZIFF} & 33,79 & 24,01 & 23,60 & 24,66 & 24,25 & 23,14 & 20,33 & 23,40 & 21,64 & 22,98 & 25,10 & \textbf{180.879} \\
		\textbf{ALL} & 33,66 & 22,81 & --    & 23,39 & --    & 22,67 & --    & 23,23 & 21,34 & 22,80 & 25,98 & \textbf{1.055.391} \\
		\hline
		\textbf{WL} & 15,06 & 4,13  & --    & 4,44  & --    & 2,90  & 1,43  & 1,30  & 1,11  & 1,39  & 6,94  & \textbf{45.867} \\
		\textbf{ENG} & 35,21 & --    & --    & --   & --    & 5,52  & 2,17  & 0,55  & 0,55  & 0,55  & 3,73  & \textbf{102.400} \\
		\textbf{EINS} & 30,14 & 0,97  & --    & 0,98  & --    & 0,27  & 0,07  & 0,07  & 0,07  & 0,07  & 5,17  & \textbf{456.667} \\
		\hline
	\end{tabular}%
	\label{tab:ratios}%
\end{table}%

In Table~\ref{tab:ratios}, we compare the compression ratios obtained. We can see that \dvtv\ is able to improve the results of \vtvdc\ (and \vtvdch) in all datasets (results with '--' indicate failed runs). This is remarkable since we are sending new words in plain form, while the best values or \vtvdc\ are drawn when it encodes the vocabulary of words with \lzma. As expected, by using not only words in the vocabulary of symbols allows \dvtv\ to overcome the original \detdc\ by more than $10$ percentage points in regular English datasets, and completely blows \detdc\ out in repetitive collections. On regular texts, \dvtv\ and \psevenzip\ obtain similar values on the largest dataset, yet in the other datasets the fact of exploiting char- rather than word-based regularities benefit \psevenzip, \lzma, and \repair. 
In repetitive text collections, char-level repetitiveness  is higher than at word-level, and in addition, the fact of sending words in plain form harms \dvtv\ compression. In practice, even though compression is good in \dvtv, it is typically far from \repair, \psevenzip, and \lzma.

In Table~\ref{tab:times}, we include both compression and decompression times. \dvtv\ is faster at compression than \psevenzip, \lzma, and \repair. It is on a par with \vtvdch, and it is slower than \bzip. Of course \detdc, which has not to deal with the detection of seen subsequences, is much simpler and faster than \dvtv.

At decompression, we can see that again \dvtv\ is the fastest technique in all cases, with the exception of \detdc\ and \vtvdc\ when dealing with non-repetitive English texts. Note that, in this case, \vtvdc\ compression is similar to that of \dvtv\ and consequently both decode approximately the same number of codewords. However, \vtvdc\ has not to perform an {\em update} procedure after decoding each symbol nor to generate a new non-terminal. \detdc\ has to decode more symbols than \dvtv\ due to its worse compression. Yet, again it is simpler because it does not have to deal with non-terminals, only with words. In the repetitive collections \dvtv\ compresses much more than \vtvdc\ and \detdc, which leads to a compressed file with much less codewords than those of \detdc\ and \vtvdc,  and this amortizes the cost of the {\em update} procedure required after decoding each codeword.

In Table~\ref{tab:memory}, we can see memory usage at compression time. In this case, our current implementation of the trie in \dvtv\ requires lots of memory. At decompression time, we only have to deal with the {\em codebook} (the size of $top$ is negligible), and the memory usage becomes much more reasonable. Yet, the number of entries in the {\em codebook} is still very high in most datasets: \{1,6M@FT91\}; \{4,3M@CR\}; \{15,4@ZIFF\}; \{81,2M@ALL\}; \{0,44M@WL\}; \{1,8M@ENG\}; \{0,3M@EINS\}.

\begin{table}[htbp]
	\centering
	\caption{Compression and decompression times (in seconds)}
	\scriptsize	
	\setlength\tabcolsep{4.5pt} 
	\begin{tabular}{c|r|r|rr|rr|r|r|r|r|r|r|}
		\cline{2-13}    \multirow{2}[4]{*}{} & \multicolumn{1}{c|}{\multirow{2}[4]{*}{Text}} & \multicolumn{1}{c|}{\multirow{2}[4]{*}{Detdc}} & v2vdc & v2vdcH & v2vdc & v2vdcH & \multicolumn{1}{c|}{\multirow{2}[4]{*}{D-v2v}} & Repair & \multicolumn{2}{c|}{lzma} & \multicolumn{1}{c|}{\multirow{2}[4]{*}{p7zip}} & bzip2 \\
		\cline{4-7}\cline{9-11}\cline{13-13}          &       &       & \multicolumn{2}{c|}{\textit{lzma  words}} & \multicolumn{2}{c|}{\textit{plain words}} &       & \textit{+sHuff} & \textit{def} & \textit{-9 -e} &       & \textit{def} \\
		\hline
		\hline
		\multirow{7}[4]{*}{\begin{sideways}Compr. time \end{sideways}} & \textbf{FT91} & 0.15  & 1.31  & 2.26  & 1.28  & 2.27  & 5.81  & 8.37  & 9.17  & 10.66 & 9.06  & 1.19 \\
		& \textbf{CR} & 0.53  & 5.77  & 19.01 & 5.72  & 19.10 & 19.03 & 39.84 & 32.61 & 44.09 & 33.67 & 4.07 \\
		& \textbf{ZIFF} & 2.12  & 32.08 & 257.42 & 31.94 & 258.03 & 86.13 & 271.02 & 120.92 & 177.86 & 128.99 & 14.52 \\
		& \textbf{ALL} & 13.25 & 292.39 & --    & 289.05 & --    & 573.09 & --    & 711.23 & 1167.86 & 768.22 & 86.71 \\
		\cline{2-13}          & \textbf{WL} & 0.48  & 17.93 & --    & 18.05 & --    & 2.96  & 14.99 & 8.88  & 23.60 & 6.23  & 2.45 \\
		& \textbf{ENG} & 1.71  & --    & --    & --    & --    & 13.39 & 58.97 & 28.37 & 57.34 & 28.31 & 8.45 \\
		& \textbf{EINS} & 6.62  & 33205.00 & --    & 33197.00 & --    & 30.48 & 205.33 & 60.98 & 115.12 & 57.13 & 54.95 \\
		\hline
		\hline
		
		\multirow{7}[4]{*}{\begin{sideways}Decompr. time\end{sideways}} & \textbf{FT91} & 0.09  & 0.08  & 0.09  & 0.06  & 0.06  & 0.09  & 0.15  & 0.18  & 0.17  & 0.19  & 0.47 \\
		& \textbf{CR} & 0.28  & 0.22  & 0.23  & 0.20  & 0.21  & 0.31  & 0.65  & 0.54  & 0.54  & 0.54  & 1.54 \\
		& \textbf{ZIFF} & 1.24  & 1.01  & 0.90  & 0.94  & 0.86  & 1.50  & 2.69  & 2.13  & 2.14  & 2.14  & 5.83 \\
		& \textbf{ALL} & 7.68  & 9.13  & --    & 8.99  & --    & 11.63 & --    & 12.13 & 12.25 & 12.15 & 33.54 \\
		\cline{2-13}          & \textbf{WL} & 0.16  & 0.06  & --    & 0.06  & --    & 0.02  & 0.28  & 0.05  & 0.05  & 0.09  & 1.00 \\
		& \textbf{ENG} & 0.85  & --    & --    & --    & --    & 0.14  & 2.50  & 0.06  & 0.04  & 0.18  & 4.08 \\
		& \textbf{EINS} & 3.22  & 0.33  & --    & 0.24  & --    & 0.05  & 1.71  & 0.15  & 0.15  & 0.71  & 9.40 \\
		\cline{2-13}    \end{tabular}%
	\label{tab:times}%
\end{table}%

\begin{table}[htbp]
	\centering
	\caption{Memory usage (in MiB) at compression and decompression}
	\scriptsize	
	\setlength\tabcolsep{4.5pt} 
	\begin{tabular}{c|r|r|rr|rr|r|r|r|r|r|r|}
		\cline{2-13}          & \multicolumn{1}{c|}{\multirow{2}[4]{*}{Text}} & \multicolumn{1}{c|}{\multirow{2}[4]{*}{Detdc}} & v2vdc & v2vdcH & v2vdc & v2vdcH & \multicolumn{1}{c|}{\multirow{2}[4]{*}{D-v2v}} & Repair & \multicolumn{2}{c|}{lzma} & \multicolumn{1}{c|}{\multirow{2}[4]{*}{p7zip}} & bzip2 \\
		\cline{4-7}\cline{9-11}\cline{13-13}          &       &       & \multicolumn{2}{c|}{\textit{lzma  words}} & \multicolumn{2}{c|}{\textit{plain words}} &       & \textit{+sHuff} & \textit{def} & \textit{-9 -e} &       & \textit{def} \\
		\hline\hline
		\multirow{7}[4]{*}{\begin{sideways}Compressor\end{sideways}} & \textbf{FT91} & 24    & 52    & 52    & 52    & 52    & 1,194 & 380   & 94    & 192   & 165   & 7 \\
		& \textbf{CR} & 53    & 157   & 157   & 157   & 157   & 2,635 & 1,286 & 94    & 504   & 193   & 7 \\
		& \textbf{ZIFF} & 126   & 625   & 625   & 625   & 625   & 10,509 & 4,585 & 94    & 674   & 193   & 7 \\
		& \textbf{ALL} & 207   & 3,509 & --    & 3,509 & --    & 46,160 & --    & 94    & 673   & 193   & 8 \\
		\cline{2-13}          & \textbf{WL} & 49    & 255   & --    & 255   & --    & 478   & 1,268 & 94    & 469   & 193   & 7 \\
		& \textbf{ENG} & 85    & --    & --    & --    & --    & 1,953 & 2,512 & 94    & 674   & 193   & 7 \\
		& \textbf{EINS} & 152   & 44,821 & --    & 9,851 & --    & 6,521 & 10,859 & 94    & 674   & 193   & 8 \\
		\hline
		\hline
		\multirow{7}[4]{*}{\begin{sideways}Decompressor\end{sideways}} & \textbf{FT91} & 4     & 10    & 20    & 20    & 10    & 20    & 13    & 9     & 15    & 17    & 4 \\
		& \textbf{CR} & 6     & 65    & 65    & 65    & 65    & 51    & 30    & 9     & 50    & 19    & 4 \\
		& \textbf{ZIFF} & 14    & 121   & 119   & 121   & 119   & 177   & 79    & 9     & 65    & 19    & 4 \\
		& \textbf{ALL} & 57    & 378   & --    & 378   & --    & 931   & --    & 9     & 65    & 19    & 4 \\
		\cline{2-13}          & \textbf{WL} & 5     & 52    & --    & 51    & --    & 6     & 11    & 9     & 46    & 18    & 4 \\
		& \textbf{ENG} & 9     & --    & --    & --    & --    & 23    & 31    & 9     & 65    & 18    & 4 \\
		& \textbf{EINS} & 14    & 73.81 & --    & 73.81 & --    & 5     & 5     & 9     & 65    & 18    & 4 \\
		\cline{2-13}    \end{tabular}%
	\label{tab:memory}%
	\vspace{-0.5cm}
\end{table}%


\section{Conclusions and future work}
We have described \dvtv, the first word-based dynamic variable-to-variable text compressor. We showed that \dvtv\ obtains competitive compression ratios (similar to \psevenzip) in English texts and that it is fast at both compression and (mainly) decompression. Even not included in the paper, note that looking for the occurrences of a given word $P$ would be possible by counting the number of escape codewords until the first occurrence of $P$ (that counter indicates the initial entry of the {\em codebook} where we add $P$). From there on, by simulating the decompressing process we need to track the occurrences of the codeword corresponding to the terminal $P$ and those codewords corresponding to all the non-terminals which include $P$. 

The main drawback of \dvtv\ is that it needs lots of memory at compression to handle the subsequences (non-terminals) in the trie. As future work, we will improve the current implementation of the trie to reduce memory requirements. We also want to apply the ideas in \cite{BFNP.tois10} to create an asymmetric lightweight version of \dvtv. This should reduce the work done by the receiver and its memory usage. In addition, the codeword associated to a given symbol $S_i$ would not vary so often, which would allow us to implement efficient direct searches for a pattern within the compressed text.


\Section{References} 
\bibliographystyle{IEEEbib}
\bibliography{biblio}

\end{document}